# Computable Contracts in the Financial Services Industry

Vinay K. CHAUDHRI[a,1]

a

**Abstract.** A computable contract is a contract that a computer can read, understand and execute. The financial services industry makes extensive use of contracts, for example, mortgage agreements, derivatives contracts, arbitration agreements, etc. Most of these contracts exist as text documents, making it difficult to automatically query, execute and analyze them. In this vision paper, we argue that the use of computable contracts in the financial services industry will lead to substantial improvements in customer experience, reductions in the cost of doing legal transactions, make it easier to respond to changing laws, and provide a much better framework for making decisions impacted by contracts. Using a simple payment agreement, we illustrate a Contract Definition Language, sketch several use cases and discuss their benefits to the financial services industry.

## 1. Introduction

A contract is an agreement between parties that governs their behavior, and often, has a legally binding effect. There has been extensive work to support contract automation as evidenced by the existence of an *e-contracts* standard [1]. Nick Szabo defined a *smart contract as* a *set of promises specified in digital form including protocols on which the parties perform on those promises* [2]. Our focus in this paper is to refine a previously articulated vision of a *computable contract* [3] that defined a computable contract *as a contract that a computer can read, understand and execute* [3]. In addition to giving a concrete example of a computable contract, we highlight how some of its capabilities are more general than available in e-contracts and smart contracts.

When we say that a computer can read and understand a contract, we mean that it has a conceptual understanding of the contract in the same way as a human does. A human who understands a contract can analyze it with the goal of negotiating better terms, assess the impact of its terms on related business operations, and identify problematic situations that might be encountered during its lifecycle. Humans also form opinions about the fairness and the risk of a contract. We envision a computable contract to exhibit similar behaviors.

Computer scientists have defined a variety of operational definitions such as a computer program's ability to answer questions and its ability to execute a contract as surrogates for human understanding. A computable contract generalizes the notion of e-contracts and smart contracts by addressing the full life cycle of a contract including

---

creation, negotiation, approval, execution, verification, analysis and compliance. Computable contracts also make it possible to analyze how a change in law or contract terms may affect the overall business. A computable contract is agnostic about the platform on which it executes. For example, it can execute on a *blockchain* [4], but it could also execute on other systems.

We begin by considering a concrete example of a financial contract that sets up automatic payment agreement for an auto finance account. We show snippets of a sample representation of the contract and illustrate how it can improve customer understanding of the contract through answering of questions and by animating and visualizing the execution of the contract over its life cycle. We then consider several other use cases where computable representation can be useful, for example, in contract libraries, interoperability, verification, etc. We conclude the paper by considering most relevant work related to computable contracts. The example considered in the paper does not cover the full lifecycle of a contract. Developing use cases that show the use of a computable contract in other phases of its life cycle, for example, contract authoring and/or verification, remains open for future work.

## 2. Automatic Payment Agreement

We consider an automatic payment agreement that a customer may sign with a financial institution when they purchase an automobile on a loan or a lease. Many financial institutions already provide online versions of such agreements. The commonplace nature of this type of agreement makes it easy to understand the difference between the current state of the art and the new capabilities we envision. The agreement consists of two parts: first, a form that includes information about the customer's auto account, address, and the bank account for payment, and second, a set of terms and conditions. We omit the form portion of the contact but show a representative set of terms of agreement in Table 1.

The terms and conditions of an automated payment agreement are a part of a larger finance agreement. The payment agreement authorizes the financial institution to make regular withdrawals from the customer's bank account. It specifies the mechanics of applying those withdrawals and deals with situations where a payment may bounce. Finally, it specifies the procedure that either party can use for canceling or terminating the agreement.

If this agreement were to be completed using a paper form, the customer would have to either go into the branch office of the financial institution or send the agreement by mail. The back office of the bank will process the agreement. The results of the processing (success or any problems that are encountered) will be sent to the customer by postal mail or electronic mail. With the computable version of the agreement, the processing that previously required work at the back office can be completed online. The customer gets immediate feedback on whether their account is current, and the date their first payment will be due under the new agreement.

## 3. Computable Representation

We propose the use of basic logic programming [5] as a Contract Definition Language. The primary constructs include *datasets, view definitions,* and *transition rules.* We illustrate these features through a few examples.

Table 1: A sample Automatic Payment Agreement

***Terms, Conditions, and Agreements***
I am voluntarily entering into this Automatic Payment Agreement ("Agreement") with the financial institution ("Institution"). I will continue to make my monthly payment until I receive written notification from the Institution that this Agreement has been processed. I will receive a confirmation letter within 10 business days from the date the Institution receives my request. The confirmation letter will provide the effective date of this Agreement. If this is a new Agreement, I understand and agree that I must continue to make monthly payments until the effective date provided in my confirmation letter. I understand that it may take up to two months for the Automatic Payment to begin. I understand that the Agreement will be effective as of the payment due date as agreed and determined under my loan agreement or as per the other due date agreed between you and the Institution.

  I understand that the Automatic Payment will occur on the payment due date or next business day, if the due date is on a Sunday or holiday. If my payment due date is on the 29th, 30th, or 31st in a month that does not have those calendar days, my Automatic Payment will occur on the last calendar day of the month. I will receive notice that the Automatic Payment occurred and the amount of the Automatic Payment on my monthly billing statement.

  I understand that any amount that I authorize through this Agreement that is greater than the Total amount due shown on my monthly billing statement will be applied to my principal balance.

  If applicable, I understand and authorize the Institution to process the final payment owed under the terms of my account in an amount less than the amount authorized under this Agreement. The amount of my final payment will be the total amount due shown on my monthly statement.

  If, during the term of this Agreement, the Institution receives a Notice of Change related to a change in my account or my bank, the Institution is authorized to update its records and continue to process my Automatic Payment. If my monthly payment amount has changed, I authorize the Institution to adjust my Automatic Payment and notify me accordingly.

  This Agreement may be canceled by the Institution for any one or more of the following reasons, and notification will be mailed to me, if: (1) the account becomes delinquent; (2) the Institution is unable to complete the Automatic Payment; (3) funds are not available at the time of the Automatic Payment.

  I understand and agree that if an Automatic Payment is returned unpaid, the Institution may attempt to transfer funds a second time. I understand that my bank may charge a fee each time the Automatic Payment is returned. The Institution may charge a returned payment fee if my bank ultimately does not honor the Automatic Payment. If the Institution does not receive the current payment due by the payment due date shown on my monthly billing statement, I may also be assessed a late charge as agreed in my loan agreement.

  I understand that if I wish to cancel my Agreement, I will notify the Institution by telephone, fax, or mail at least three business days prior to the next scheduled payment due date, otherwise the Automatic Payment will occur as previously agreed. I also understand that I may provide verbal instructions to update my Automatic Payment by calling the number below.

*3.1. Dataset*

The agreement begins by introducing three kinds of objects as defined below:
    *class(financial_institution)*
    *class(customer)*
    *class(automatic_payment_agreement)*

To capture the remaining elements of the first sentence, we also need to refer to the customer account information that was entered in the form using the following facts:
    *class(auto_finance_account)*
    *instance_of(afa,auto_finance_account)*
    *instance_of(apa,automatic_payment_agreement)*
    *has_apa(afa, apa)*
    *instance_of(bank_1, financial_institution)*
    *has_permission(bank_1, apa)*
    *has_pretty_name(mapa, "Automatic Payment")*

In the above dataset, we use the special relations **class**, **subclass_of**, and **instance_of** to capture the ontological information. The facts in the above dataset capture that the customer is granting a permission to the financial institution as per the automatic payment agreement **apa** for the auto finance account number **afa**.

*3.2. View Definitions*

For capturing some of the terms and conditions, we need to write rules. In basic logic programming, we refer to such rules as view definitions. As an example, we can specify the second sentence as the following view on a customer's obligations.

*has_obligation(c,make_payment) :-*
    *instance_of(c,customer) & ~existing_apa &  today(DD,MM,YY) &*
    *new_apa_from(DD1,MM1,YY1) & date_before(DD,MM,YY,DD1,MM1,YY1)*

The above rule calculates one of the obligations of the customer. The customer has an obligation to keep making payments until the date the new agreement goes into effect. In this rule, the predicate *existing_apa* checks if the customer already has a prior automatic agreement in place.  We perform this check using the following rule:

*existing_apa :-*
    *has_apa(afa,APA) & instance_of(APA,automatic_payment_agreement)*

In the above rule, the predicates **has_apa** performs a look up in the database using external calls to retrieve the account number and any prior agreements.

*3.3. Dynamic Rules*

We use dynamic rules whenever we need to make changes to the dataset. Changes are natural as the contract executes over time. One example of such change is different states that the contract may go through: **active, invoiced, payment pending, overdue** and **terminated**. These states define a finite state automaton. The following dynamic rule illustrates how we can capture transitions between states.

*today(M,D,Y) & mp1(M,MP1) & yp1(M,Y,YP1) &*
*~invoiced(M,Y) & has_invoice_day(A,I) & ~payment_received &*
*has_termination_date(afa,MM,DD,YY) & date_before(M,D,Y,MM,DD,YY) &*
*current_balance(A,B) & monthly_payment(A,C) & evaluate(minus(B,C),B1) & ==>*
       *~today(M,D,Y) & today(MP1,I,YP1) & current_balance(A,B1) &*
       *~current_balance(A,B) & invoiced(MP1,YP1)*

A dynamic rule has the general form of **condition ==> effects**. In the above dynamic rule, predicates **mp1** and **yp1** calculate the next month and the year of a given date. We check to make sure that we have not already invoiced the account for that month, and that the current date is before the termination date of the agreement. If the conditions are satisfied, we issue the invoice, and update the outstanding balance to reflect the pending invoice. Just like evaluating view definitions, the dynamic rules must connect to the database management system of the financial institution for execution.

     The representation of state as dynamic rules is more general than capturing the contract as a finite state automaton [6]. The representation of state in logic programming also includes datasets and view definitions, which can share predicates with dynamic rules, thus, providing a comprehensive framework to capture both static and dynamic aspects of a contract.

## 4. Using Computable Representation

We will now consider answering questions and simulating the execution of the contract. We assume the existence of a rule engine capable of necessary processing [7].

*4.1. Answering Questions*

We extend the presentation of the terms and conditions with a set of frequently asked questions (FAQ) that a user poses simply by a click. The users often do have questions about the terms and conditions, but they are often intimidated by the impenetrability of the legal language. We can use the computable representation as a basis to automatically generate a set of FAQs. We show below a few sample FAQs.

| Ask | Who is authorized to accept this agreement? |
| Ask | What permissions am I granting as part of this agreement? |
| Ask | What obligations do I have under this contract? |
| Ask | When will be my first payment under this new agreement? |
| Ask | On what date will the withdrawals be applied to my account? |

We can answer these questions automatically by posing queries against the computable representation. The answer to some of these questions follow directly from a lookup on the dataset of the contract, but for other questions, we must derive the answer. For example, to answer the second question, we need to compute the **has_permission** view, to produce the answer as shown below.

| Ask | What permissions am I granting as part of this agreement? |

"Setup Automatic Payment"
"Make Monthly Withdrawal"

For answering some questions, we need to look outside the representation of the contract. For example, for the fourth question, we must look up the customer account, determine by when the new agreement will be in place, and provide an answer accordingly. We show a sample answer next.

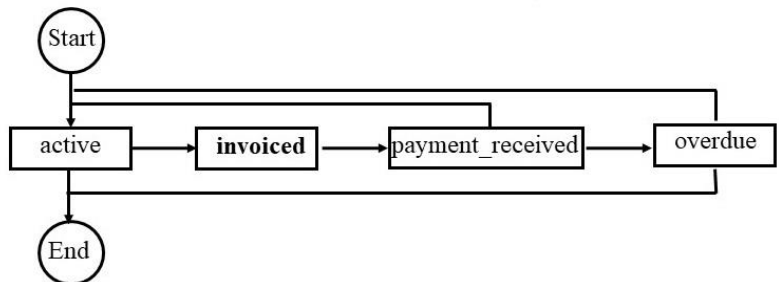

We can cross-link the computed answers to the relevant portions of the terms and conditions to provide provenance for the answer. We can ensure that the generated set of FAQs leverage each term and condition in at least one question.

*4.2. Simulating the Execution*

Using the dynamic rules, we can also show a visualization of the different states that a contract might go through over its lifetime. When the customer initially signs the agreement, the contract is in an *active* state, and the current account balance and pending withdrawals are both zero. We can simulate the execution by clicking on the *Advance Time* button. Using the dynamic rule that we showed earlier, the system enters in the *invoiced* state, and the current balance is $500 assuming that is the monthly payment. We show this state in the diagram below.

| Current Month: 2 | Current Balance: 500 | Pending Withdrawal: |
|---|---|---|
| Advance Time | | Payment Received |

As the receipt of payment is an external event, we can use the dynamic rules to update the dataset in response to such external events. For the simulation considered here, we indicate external events by clicking on the **Payment Received** button. If the financial institution does not receive the payment in time, the account goes into an ***overdue*** state, because of which it can terminate the account.

The state diagram we have considered here is simple. In general, the state diagrams for contracts can be very complex. With a facility to visualize different ways the contracts might unfold; a user will gain much deeper understanding of the contract.

**5. Use Cases for Computable Representation**

Answering questions about a contract and simulating its execution are useful for improving customer understanding of the contract. In this section, we consider several other situations where a computable representation has business value.

*5.1. Better specification of the contract*

The process of creating a computable representation of a contract reveals ambiguities and errors in a contract. Addressing those leads to a better specification of the contract. As an example, the last sentence of the first paragraph of the sample contract in Table 1 says that the due date can also be any *other due date agreed upon by you and Bank*. The notion of *other due date* in this clause is not available in terms and conditions. For a computable representation, we need a clear definition for what other places we should check for any agreements regarding the due date. This example illustrates that creating a computable representation forces us to specify the contract more clearly and precisely.

*5.2. Contracts of the Future*

We can envision a future in which we specify most of the contract in a contract definition language. We argue that the existing contracts already have clauses that are amenable to such specification. For example, we can specify the second sentence of the second paragraph as follows:

*adjusted_due_day(D) :-*
 *has_due_day(A,29) & this_month(M) &*
 *this_year(Y) & last_day(M,Y,D) & distinct(D,29)*

Some clauses require human intervention, for example, in the sixth paragraph the conditions under which the financial institution may cancel the contract are spelled out, but such cancellation is not automatic. As the decision must be made by a human, such clauses cannot be fully automated.

*5.3. Verification and Compliance*

In the current practice, once the language of the contract is in place, we give it to the programmers to implement it in code. We must then take steps to ensure that the code faithfully implements the intent of the contract. For a computable contract, we can write the contract as a logic program. The customers can understand and agree to such a contract by asking questions and simulating its execution. Once they agree to the computable contract, the same logic program could execute within the computer systems of the financial institution. The end-user directly verifies the contract, because the version of the contract they sign, is the version that executes. Realizing this vision requires substantial new research [8].

*5.4. Contract Interoperability*

Financial institutions routinely exchange contracts in PDF format in a way that the structured data is lost. A computable contract enables businesses to exchange the structured data of a contract as well as the rules and the logic of the contract. As an example, sometimes, the financial institutions have affiliates, can be bought over, or sell their loans. In such cases, the financial institution can transmit the computable version of a contract, with the expectation that the receiving party will execute it without much manual processing or human intervention. Such interoperability has the potential to lead to substantial savings for both institutions.

*5.5. Contract Libraries*

Even though organizations have started to use contract templates to gain efficiency in authoring, once the contract authoring is complete, the contract lives as a PDF document. We argue that the structure created during the authoring phase of a contract must persist over the life cycle of the contract. In the world of computable contracts, the contract lives as a logic program throughout its life. If a financial institution consolidates all its computable contracts into one library, it enables a new class of analyses that are not possible today. For example, a change in interest rate may affect thousands of contracts, and it is a painful process to change all the agreements. As a result, the financial services industry uses patchwork solutions, i.e., the contract is not changed, but an addendum to the contracts affects the desired change. In a computable contract, such a change is much more easily achievable. More broadly, we can ask new kinds of questions as follows. How a change in law would affect different contracts that an institution has? If a financial institution were to change some policy (e.g., interest rate), what impact might it have on its overall business?

**6. Related Work**

We will now consider early work on smart contracts, some recent contract automation in the industry, and an important class of contracts involving securities and derivatives transactions. An extensive survey of computable contracts exists elsewhere [9].

*6.1. Smart Contracts*

In 1996, Nick Szabo defined a smart contract as "a set of ***promises***, specified in digital form, including ***protocols*** using which the parties perform on those promises" [2]. The set of ***promises*** in a smart contract can be contractual or non-contractual. A smart contract consists of lines of code as well as software that prescribes its conditions and outcomes. A smart contract embeds contractual clauses and conditions within code as software. By ***protocol***, we mean an algorithm or a set of rules that define how an organization must process the smart contract. The primary use case for smart contracts is process automation. The business value of process automation is easy to justify. Therefore, most work on smart contracts has focused on those clauses that are amenable to automation. There is recognition that the smart contracts exist on a spectrum. On one end, the contract exists entirely in code, and on the other, the contract exists in natural language with automatable portions available as code. Many current efforts on smart contracts use block chain as the platform for executing a smart contract.

      Computable contracts and smart contracts share the core idea that a contract exists as code that we can use for automated execution. Our view of computable contracts goes far beyond automated execution. We envision using the computable representation for asking questions, simulating the execution, verification, interoperability, etc. We have also implicitly argued in favor of a declarative language, such as logic programming, as a preferred contract definition language for computable contracts. We have shown that logic programming provides a framework in which we can reason with the static properties of a contract (i.e., answering questions), and handle the dynamics aspects (i.e., using dynamic rules).

*6.2. Recent Deployments*

Large financial institutions enter millions of contracts every year, which fall into several thousand templates. Therefore, using a template-based contract authoring is a natural first step in the direction of computable contracts. Hot Docs™ is an example of such a product (See https://www.hotdocs.com/), which enables creation of contracts from templates. It provides a rule language to capture conditional branches through a contract and is used by many large financial organizations for drafting contracts.

DocuSign®, a provider of electronic signature platform (see https://www.docusign.com), has introduced *Smart Clauses*®, which is a programmed element of a traditional or a new contract. Smart Clauses® use technology developed as part of the research project Accord (See https://www.accordproject.org/). *Smart Clauses*® realize only a small aspect of the computable contracts vision.

*6.3. Swaps and Derivatives Contracts*

A high-value derivatives transaction establishes a financial relationship between counterparties that may last many decades and may involve a very substantial notional sum. This requires extensive legal protection. Many derivatives transactions utilize legal agreements based on standardized legal documentation provided by the International Swaps and Derivatives Association (ISDA) (See http://www.isda.org).

ISDA is an independent standards body and is actively engaged in leveraging the computable contracts technology [10]. Towards that end, ISDA has released a contract-authoring tool called ISDA Create [11]. ISDA Create allows firms to produce, deliver, negotiate and execute contracts and capture, process and store data from such contracts. It allows standard elections for contract terms, but it also allows firms to customize on a party-by-party basis. There are no restrictions on what parties may agree on a bilateral basis on the platform.

Standardization of clauses in ISDA Create reduces variation among contracts, enables their computer understanding, and supports automation. ISDA has also released a common data model for this tool that specifies the data used in ISDA Create. We need significant additional research to specify the code that automates the processes and to ensure that the code complies with the contract [8]. Multiple formalization approaches are being pursued and their relative tradeoffs need to be better understood [12-17].

**7. Summary**

In this paper, we refined the prior vision of computable contract which defines it as a contract that a computer can read, understand and execute. We provided details on what it means for a program to read, understand and execute a contract as abilities to answer questions as well to execute a contract. A computable contract is more general than a smart contract in that the use of computable representation is not limited to automatically executing some aspect of a contract. We can use a computable contract for answering questions, simulating execution, and for doing verification and analysis. A computable contract covers the full life cycle of a contract from authoring, negotiation, implementation, and eventual execution and termination. We took an example of a simple automatic payment agreement and highlighted how it advances the current generation of online contracts. Computable contracts can improve user understanding of the contract and serve as a basis for verifying that the implementation complies the contract. Even though we are starting to see adoption of rudimentary forms of computable contracts, significant untapped potential exists. Realizing such potential requires research on formal contract specification languages, and reasoning and analysis tools. We need to make the benefits of computable contracts more concrete through creation of high value use cases for financial services industry [18]. Ultimate adoption of this technology, however, will happen when several financial institutions start producing computable contracts that they share with each other so that everyone can exploit the structured information in them to their individual benefits. Such individual benefits will forward an incentive system that will encourage a large-scale and systemic move towards computable contracts.